\documentclass[journal,a4paper]{IEEEtran_nssmic}

\usepackage{cite}      

\usepackage{graphicx}  

\usepackage{subfigure} 

\let \IG \includegraphics

\usepackage{xspace}

\newcommand{\sa}{superattenuator}

\newcommand{\VIRGO}{\textsc{VIRGO}\xspace}

\newcommand{\FP}{Fabry-Perot}

\newcommand{\prc}{power-re\-cycling cavity}

\newcommand{\pom}{power-recycling mirror}

\newcommand{\Mi}{Michelson interferometer}

\newcommand{\mc}{mode cleaner}
\newcommand{\mHc}{mode-cleaner}
\newcommand{\omc}{output mode cleaner}
\newcommand{\imc}{input mode cleaner}

\newcommand{\gw}{gravitational wave}

\newcommand{\gHw}{gravi\-ta\-tion\-al-wave}

\newcommand{\M}[1]{\ensuremath{{\rm TEM}_{#1}}}

\newcommand{\FSR}{free spectral range}
\newcommand{\pd}{photo diode}

\newcommand{\bs}{beam splitter}

\newcommand{\mFig}[1]{Figure~\ref{#1}}

\begin{document}

\title{
Automatic Mirror Alignment for VIRGO: First experimental demonstration of the Anderson technique 
on a large-scale interferometer.}

\author{
F.~Acernese\authorrefmark{6}, 
P.~Amico\authorrefmark{10}, 
S.~Aoudia\authorrefmark{7}, 
N.~Arnaud\authorrefmark{8},
S.~Avino\authorrefmark{6},
D.~Babusci\authorrefmark{4}, 
G.~Ballardin\authorrefmark{2}, 
R.~Barill\'e\authorrefmark{2}, 
F.~Barone\authorrefmark{6}, 
L.~Barsotti\authorrefmark{11}
% ...
\\
\authorrefmark{1}Laboratoire d'Annecy-le-Vieux de Physique des Particules, Annecy-le-Vieux, France;\\
\authorrefmark{2}European Gravitational Observatory (EGO), Cascina (Pi), Italia;\\
\authorrefmark{3}INFN, Sezione di Firenze/Urbino, Sesto Fiorentino, and/or Universit\`a di Firenze, and/or Osservatorio Astrofisico di Arcetri, Firenze and/or Universit\`a di Urbino, Italia;\\
\authorrefmark{4}INFN, Laboratori Nazionali di Frascati, Frascati (Rm), Italia;\\
\authorrefmark{5}SMA, IPNL, Villeurbanne, Lyon, France;\\
\authorrefmark{6}INFN, sezione di Napoli and/or Universit\`a di Napoli "Federico II" Complesso Universitario di Monte S.Angelo, and/or Universit\`a di Salerno, Fisciano (Sa), Italia;\\
\authorrefmark{7}Departement Artemis -- Observatoire de la C\^ote d'Azur, BP 42209 06304 Nice, Cedex 4, France;\\
\authorrefmark{8}Laboratoire de l'Acc\'el\'erateur Lin\'eaire (LAL), IN2P3/CNRS-Univ.   de Paris-Sud, Orsay, France;\\
\authorrefmark{9}ESPCI, Paris, France;\\
\authorrefmark{10}INFN, Sezione di Perugia and/or Universit\`a di Perugia, Perugia, Italia;\\
\authorrefmark{11}INFN, Sezione di Pisa and/or Universit\`a di Pisa, Pisa, Italia;\\
\authorrefmark{12}INFN, Sezione di Roma and/or Universit\`a "La Sapienza",  Roma, Italia.
\thanks{Send correspondence to A. Freise, E-mail: andreas.freise@ego-gw.it}}


\maketitle

\begin{abstract}
The French-Italian interferometric gravitational wave
detector VIRGO is currently being commissioned. Its 
principal instrument is a Michelson laser interferometer with 3 km
long optical cavities in the arms and a power-recycling mirror.
The interferometer resides in an ultra-high vacuum system and 
the mirrors are suspended from multistage pendulums for seismic isolation.

This type of laser interferometer reaches its maximum sensitivity only
when the optical setup is held actively very accurately at a defined operating
point: control systems using the precise interferometer signals stabilise the 
longitudinal and angular positions of the optical component. 
This paper gives an overview of the control system for the angular degrees
of freedom; we present the current status of the system
and report the first experimental demonstration of the Anderson 
technique on a large-scale interferometer.
\end{abstract}

\begin{keywords}
gravitational wave detector, laser interferometer, automatic alignment, VIRGO
\end{keywords}


\section{Introduction}
\PARstart{T}{he} French-Italian collaboration VIRGO~\cite{VIRGO:prop} 
has built a large-scale interferometric \gHw\ detector
near Pisa, Italy. The main instrument is a Michelson interferometer (MI) with 3\,km long
\FP\ (FP) cavities in its arms. 
High-quality optics are suspended to act as quasi-free
test masses at the end of the \Mi\ arms so that a \gw, passing
perpendicular to the detector, will be detected in the 
interferometer signal. The \FP\ cavities in the arms enhance the light
power and thus increase the optical gain of the interferometer.
The apparatus is designed to achieve a relative displacement sensitivity
of better than $\delta l/l=10^{-21}\,/\sqrt{\rm Hz}$ between 20\,Hz and
10\,kHz. The sensitivity will be limited by seismic disturbances below 3\,Hz, 
by thermal noise up to 100\,Hz and by shot noise
for higher frequencies. 

In order to reach this extreme sensitivity, the \VIRGO detector uses special techniques to minimise the
coupling of noise into the interferometric signal. The large optics (mirrors
and beam splitters) are super-polished
fused silica pieces with very low absorption and scattering. They are located
in an ultra-high vacuum system and suspended from a sophisticated seismic isolation system,
the so-called \sa\ \cite{SA}. The \sa\ consists of a multistage pendulum and offers an enormous
passive isolation from seismic motion for Fourier frequencies larger than its mechanical
resonance frequencies ($10^{14}$ at 10\,Hz). However, the motion at the resonance frequencies 
(between 10\,mHz and 4\,Hz) can be large and must be reduced by active control.

The optics of the \Mi\ and of the cavities have to be well-aligned
with respect to each other and with respect to the incoming beam 
to reach their high optical qualities.
The angular degrees of freedom of the suspended optics show a 
non-negligible deviations which, if uncontrolled, distort the cavity
eigen-modes and also the interference at the \bs. 
This causes power modulation of the light fields;
long term drifts will make a longitudinal control impossible after
a certain amount time; and furthermore, 
mis-alignments increase the coupling of other noise sources 
into the main detector output. 
To guarantee a stable long-term operation and a high sensitivity the angular
degrees of freedom have to be actively controlled. 
\begin{figure*}[t]
\begin{center}
\IG [scale=.23, angle=0] {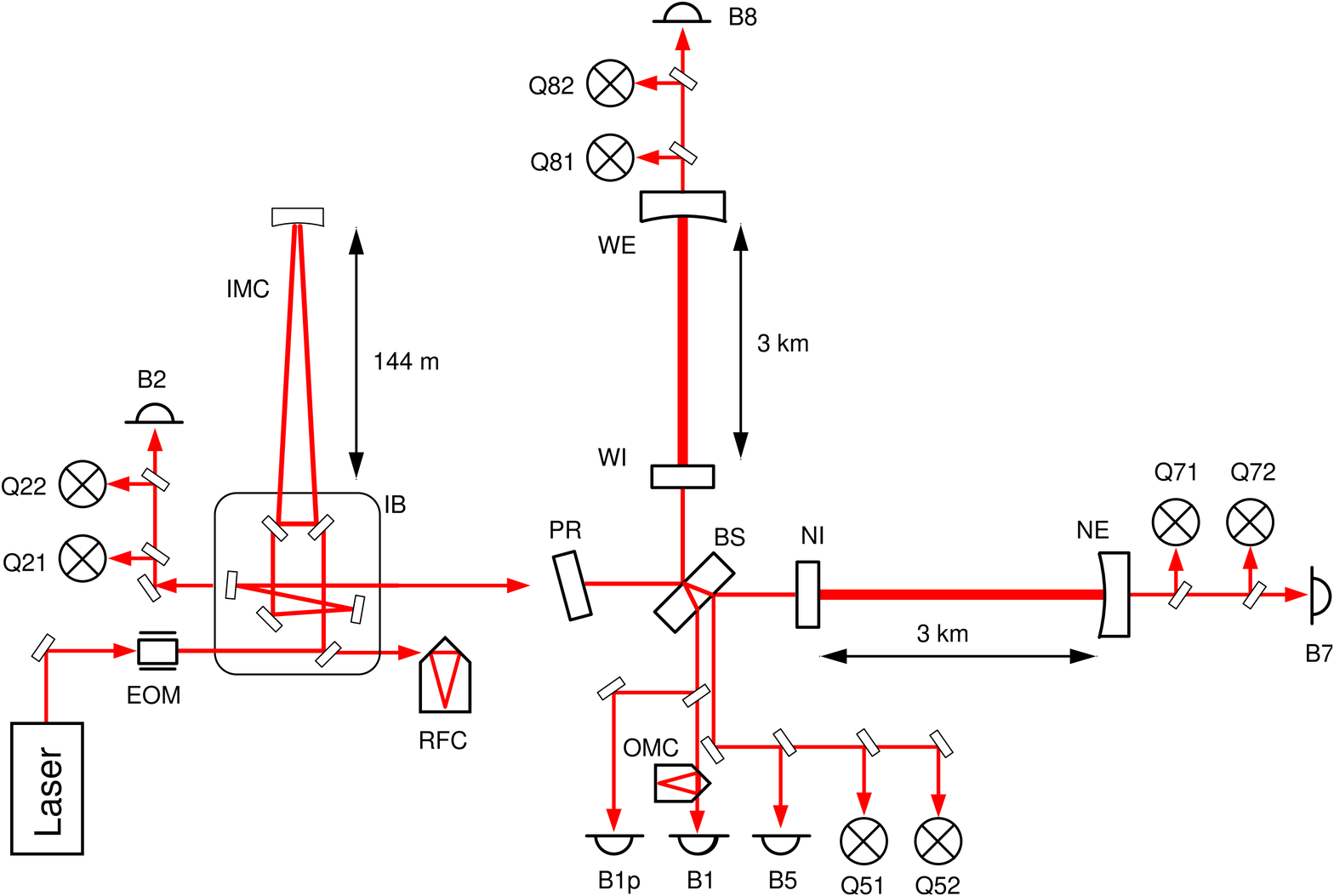}
\end{center}
\caption{\label{fig:optical-layout}A simplified schematic of the
optical design for VIRGO: The laser beam is directed on the \emph{detectors table}
(DT) into the first vacuum chamber, the injection tower, in which all
optical components are attached to a suspended optical bench, the \emph{injection
bench} (IB). After passing the \emph{input \mc} (IMC), the beam is injected
through the \emph{\pom} (PR) into the main interferometer. The beam is split and
enters the two 3\,km long arm cavities, the \emph{west arm} (WA) and 
\emph{north arm} (NA). The \Mi\ (MI) is held at the dark fringe so that
most of the light power is reflected back to the \emph{\pom} (PR). In the final configuration the PR 
together with the MI form a 
\FP-like cavity in which
the light power is enhanced. In the recombined configuration the PR is largely misaligned and can 
be simply 
considered as an attenuator with a transmission of $8\%$. 
The light from the
dark port of the beam splitter is filtered by an \emph{\omc} (OMC) before being detected on a set
of 16 photo diodes (B1), which generate the main output signal of the detector. 
The other photo diodes shown in this schematic with names starting with {\bf B}
are used for longitudinal control of the interferometer; diodes named with a {\bf
Q} represent split photo detectors used for alignment control.}
\end{figure*}

\section{Optical Layout}
\mFig{fig:optical-layout} shows a simplified optical layout of the VIRGO interferometer
in the recombined configuration. 
Although the MI in the recombined configuration does not provide
a detector sensitive enough for detecting gravitational waves, 
it is the final step on the way to the
detector with recycling. 
It allows to characterise the optical signals
to implement the proper control system for the recycled interferometer, 
for example, for the frequency stabilisation or the automatic alignment.

The laser light, 20\,W @1064nm provided by an injection-locked master-slave
solid state laser (Nd:YAG), enters the vacuum system at the 
\emph{injection bench} (IB). The beam is spatially filtered by a 144\,m long input \mHc\ cavity (IMC)
before being injected into the main interferometer. 

The laser frequency is pre-stabilised using the IMC cavity as a reference. The
low-frequency stability is achieved by an additional control system that stabilises
the IMC length below 15\,Hz to the length of a so-called reference 
cavity (RFC). 

A beam with 10\,W of power enters the Michelson interferometer through the \pom. 
In the recombined configuration the \pom\ is largely misaligned so that it
merely attenuates the beam without interacting with the rest of the interferometer.
In consequence, a beam with approximately 800\,mW is impinging on the main \bs.
It is split
into two beams that are injected into the 3\,km long arm cavities.
The finesse of the arm cavities is approximately 50; this yields a circulating light power
of 13\,W. The flat input mirrors and the spherical end mirrors 
form a stable resonator with the beam waist being at the input
mirrors with a radius of approximately 20\,mm.

The MI is held on the dark fringe, and the expected \gw\
signal will be measured in the beam from the dark port, which is 
passed through an
\emph{\omc} (OMC), a 2.5\,cm long rigid cavity and then detected on \pd\ B1
(in fact, B1 is a group of 16 InGaAs \pd s).

The interferometer control systems utilise the interferometer output signals and 
a modulation-demodulation method. For
this purpose the laser beam is modulated in phase with a electro-optic
modulator (EOM, see \mFig{fig:optical-layout}) at $f_{\rm RF}=6.26$\,MHz. Such a
phase modulation generates new frequency components 
with a frequency offset of $\pm f_{\rm RF}$ to the frequency
of the laser beam $f_0$. These frequency components are called \emph{sidebands},
and the light field at $f_0$ is called \emph{carrier}

The photo
currents detected by the photo diodes in several output ports of the
interferometer are then demodulated with the same frequency $f_{\rm RF}$
or multiples of that frequency.
The demodulated signals from single element diodes (B1 to B8) are used for the
length control of the interferometer, whereas the demodulated signals 
from split photo detectors (Q21 to Q82) provide control signals for the
angular degrees of freedom of the interferometer mirrors.


The modulation-demodulation technique is commonly used in many interferometers,
especially in the other interferometric gravitational wave detectors
(LIGO, TAMA and GEO\,600~\cite{ligo,tama,geo}). However, for each 
interferometer a unique control topology has been developed, 
depending on the details of the experimental realisation of the
optical system~\cite{Lisa}.

\section{Alignment sensing and control}
The automatic alignment of the interferometer consists
of two types of control: the \emph{linear alignment} measures
the mirror positions with respect to
the laser beam and feeds back to the mirrors angular positions
(bandwidth $\approx 5$\,Hz), and the 
\emph{drift control}: a low frequency (bandwidth $<100$\,mHz)
control measures the beam axis position in the two 
interferometer arms and feeds back to the \bs\ and the
injection bench angular position. 

The specifications given for the alignment control require
that the angular fluctuations are reduced to $10^{-7}\,{\rm rad}_{\rm RMS}$
for the recycling mirror, $2\cdot10^{-8}\,{\rm rad}_{\rm RMS}$ for the 
cavity input mirrors and $3\cdot 10^{-9}\,{\rm rad}_{\rm RMS}$ for the
cavity end mirrors. 

During the lock acquisition process the mirrors and the \bs\
are controlled by local control systems that keep the 
mirrors aligned with respect to local references\cite{LC}.
These controls can reduce the angular fluctuations of the
mirrors to a few microradians RMS over a time period of
one hour. But the local control can neither achieve the required
long term stability nor the low noise spectral density.

The purpose of the linear alignment is to provide a global control 
for the mirror alignment that uses the most 
precise error signals derived from the interferometer itself.
Thus, after the interferometer has reached its operating
condition, the controls of the angular positions should be switched 
from the local controls to the linear alignment.

The control topology for the linear alignment of \VIRGO 
has been designed and tested using time domain simulations and a table-top
prototype experiment\cite{LA}.

\subsection{Modulation frequency}
The control system for the \VIRGO interferometer
is unique in two ways: first, only one modulation is used to 
derive control signals for all longitudinal and angular degrees
of freedom of the main interferometer. Second, as the
only large-scale interferometer \VIRGO makes use of the Anderson technique~\cite{anderson}
in the sense that it uses the light transmitted by the arm cavities
to generate differential wave-front signals.

This design requires a carefully chosen modulation frequency and
carefully tuned cavity lengths. The respective cavities are: the input
\mc\ cavity of $L_{\rm IMC}\approx 144$\,m, the arm cavities with $L_{\rm FP}\approx3$\,km
and the short \prc\ formed by PR with NI and WI with a length of
$L_{\rm PRC}\approx12.07$\,m. The sideband frequency and the cavity lengths have to 
be set following an exact scheme:
\begin{itemize}
\item{in order to pass the sidebands through the IMC the modulation frequency must be an exact multiple
of the IMC's \FSR: $f_{\rm RF}~=~N~{\rm FSR}_{\rm IMC}$ with N as an integral number.
This condition has to be tuned very carefully because a 
mismatch of only a few Hz couples lengths noise of the IMC into
the main interferometer signals.}
\item{for the longitudinal control signals the sidebands should be anti-resonant
in the arm cavities and resonant in \prc. 
This can be achieved if the sideband frequency is a) not an exact multiple of the arm cavities
\FSR\ and b) the sideband frequency is chosen such that $f_{\rm RF}={\rm FSR}_{\rm PRC}/2$.
An accurate tuning of this condition is desirable to achieve an
intrinsic decoupling of the longitudinal control signals but a 
deviation of a few kHz can be tolerated. Note that in 
the recombined configuration
the \prc\ is not present so that this condition can be ignored.}
\item{the Anderson technique requires the modulation frequency  
to be a resonance frequency of the \M{01} mode in the arm cavities. Thus:
\begin{equation}
f_{\rm RF}~=~N~ {\rm FSR}_{\rm FP} + f_{\rm sep}
\end{equation}
with N as a integral number, ${\rm FSR}_{\rm FP}$ the \FSR\ of the arm cavity 
and $f_{\rm sep}$ the mode separation frequency, which is the difference between the
resonance frequencies of TEM modes for different mode numbers:
\begin{equation}
\begin{array}{ccl}
f_{\rm sep}&=&f_{\rm n+1,m}-f_{\rm n,m}\\
&=&\frac{c}{2\pi L_{\rm FP}}
\arccos{\left(\sqrt{1-L_{\rm FP}/R_C}\right)}
\end{array}
\end{equation}
with $f_{\rm n,m}$ being the resonance frequency of a \M{nm} mode, $c$, 
the speed of light and $R_C$ the radius of curvature
of the cavity end mirror.

The modulation frequency must be tuned with respect to this value given by
the cavity length and the mirror curvature. The tolerance is $\pm 500$\,Hz,
given by the linewidth of the arm cavities.}
\end{itemize}
\begin{figure}[h]
\begin{center}
\IG [scale=0.7, angle=0] {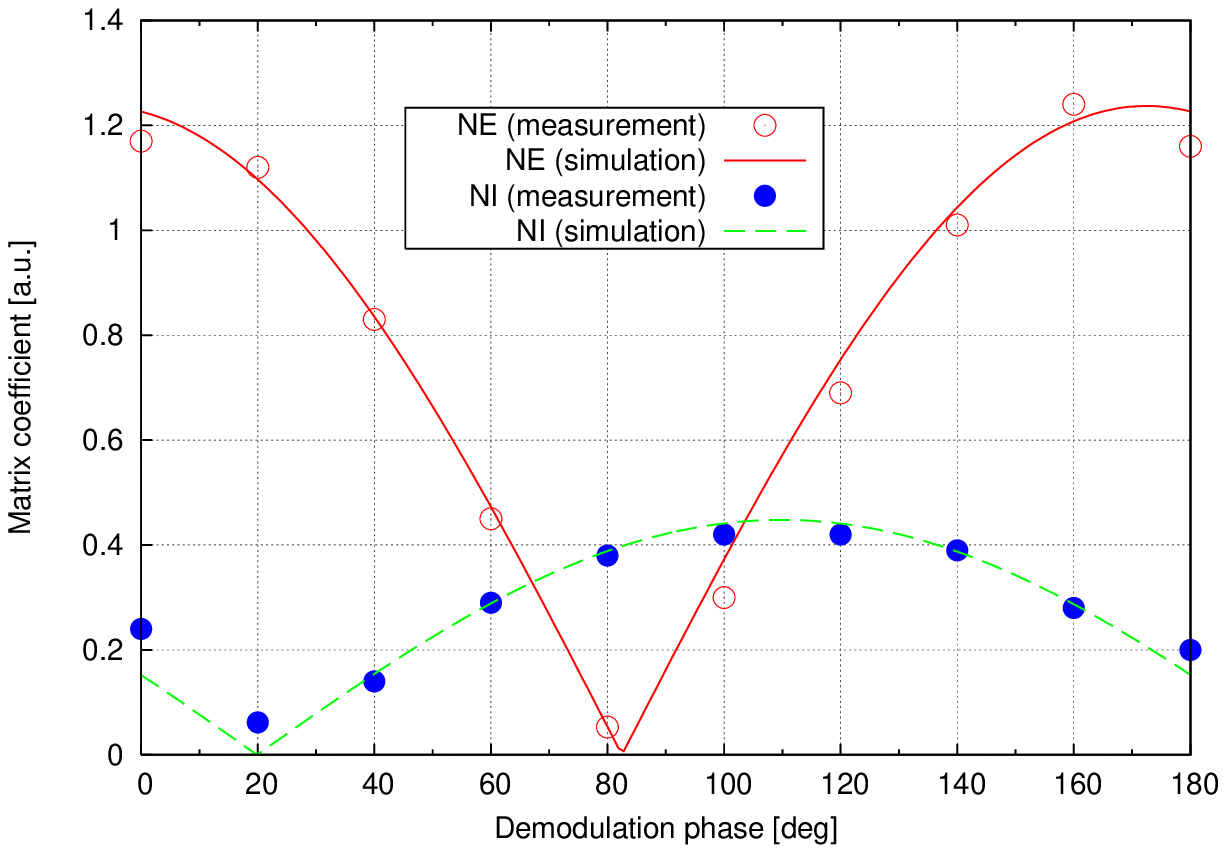}
\end{center}
\caption{\label{fig:q71tx}This plot shows the amplitude of the sensing matrix
coefficients for Q71 as a function of the electronic demodulation phase: 
the measured data were obtained with the north arm cavity 
longitudinally controlled: 
a sinusoidal disturbance of known amplitude was injected to the respective
mirror and the corresponding amplitude in the quadrant signal measured. The simulated values
were produced by computing the DC value of the transfer function: mis-alignment 
angle$\rightarrow$ differential wave front signal. By choosing the correct
demodulation phase a complete signal separation can be achieved.}
\end{figure}
 
\subsection{Alignment error signals}
The sensors for the alignment control are \emph{quadrant split
photo detectors} (or \emph{quadrant diodes} for short). These are \pd s with 4 separate elements,
so called quadrants. The quadrants have an active area of 25\,$\rm mm^2$ each
and are oriented at $45^\circ$. The four quadrants (up, down, left, right) 
are separated by a 125\,micrometers wide gap and form together
a circular photo sensitive area.

Each quadrant diode can provide a multitude of signals:
The sum over four quadrants gives the same signal as a normal
\pd. In order to generate alignment control signals the differences
between the upper and lower elements and the left and right 
elements are computed. These signals give the vertical
and horizontal positions of the beam on the diode and serve
as error signals for the drift control.

A demodulation of a differential quadrant output at $f_{\rm RF}$
yield two additional outputs: the in-phase and quadrature signal
contain information about the angle and position 
(vertical and horizontal respectively) of the
phase front of the carrier with respect to the sidebands. These signals 
can provide very accurate error signal for an automatic alignment
system if the optical setup
is composed in such a way that a misalignment of a mirror produces a 
spatial separation
between the carrier and the sidebands.
This method
is called differential wave-front sensing and is used to obtain 
error signals for the linear alignment control.
For a single
\FP\ cavity the alignment control signals can be derived in 
reflection of the cavity (Ward method \cite{ward}) or
in transmission of the cavity (Anderson method \cite{anderson}).
The automatic alignment system in \VIRGO uses also the
light transmitted by the arm cavities and can thus be considered as 
an extension of the Anderson technique:
The quadrant diodes are arranged in pairs. 
Four sets of two quadrants each are located
in four of the interferometer outputs (see \mFig{fig:optical-layout}): 
in reflection (Q21, Q22), at the reflection of the anti-reflex
coated surface of the \bs\ (Q51, Q52) and in transmission of the arm
cavities (Q71, Q72, Q81, Q82). 
Adjustable lens telescopes are used
so that the beam impinging on the two quadrants of each pair
has a Gouy phase difference of $90^\circ$.


\subsection{Sensing Matrix}
The linear alignment of the \VIRGO interferometer controls ten 
degrees of freedom: the angular positions of the four
arm cavity mirrors (NI, NE, WI and WE) plus the angular position
of the PR mirror\footnote{The angular motion of the \bs\ does not represent a separate
degree of freedom but is indistinguishable from a motion of the
WI mirror.}. 

In the recombined configuration the two arm cavities
are completely decoupled and can be treated as completely separated. 
In the case of final \VIRGO configuration this is not
longer the case; all 10 degrees of freedom are strongly
coupled. 

In general, a demodulated signal of a quadrant detector
represents a linear combination of the various angular degrees of freedom
of the optical system. In order to design  control systems 
of a complex system like the \VIRGO interferometer it is important to 
reduce the coupling between the various degrees of freedom in the sensors. 

In the following we assume that
the cavity mirrors and the optical readout have been set up well, so that 
the horizontal and vertical degrees of freedom
can be treated completely separately. 

A further separation of the several degrees of freedom in one
quadrant diode signal can be achieved by tuning the Gouy phase $\Psi$. For
a simple \FP\ cavity with one plane and one curved mirror one
can use one quadrant diode in near field ($\Psi=0^\circ$) to measure the 
alignment of the flat mirror and one quadrant in the far field ($\Psi=90^\circ$)
to measure the curved mirror misalignment. 

The optimum Gouy phases for the \VIRGO interferometer have been computed using a interferometer
simulation. Adjustable telescopes are used to set the Gouy phases to these values and
to allow a further adjustment to correct for the unavoidable discrepancies between the 
experimental realisation and the simulation.

Finally the demodulation phase for each quadrant diode can be set in order
to minimise the coupling of error signals .
\begin{figure}[h]
\begin{center}
\IG [scale=0.55, angle=0] {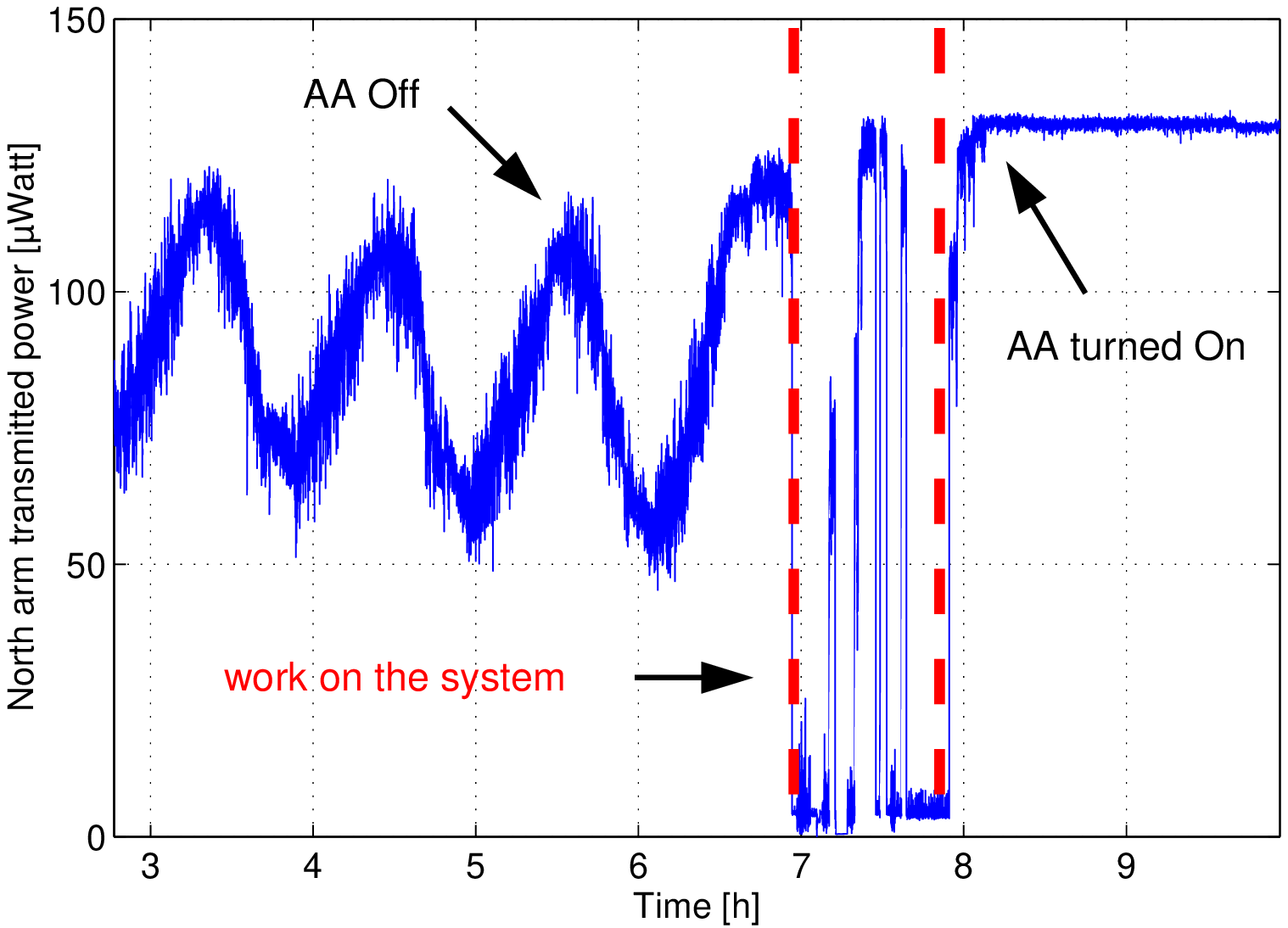}
\end{center}
\caption{\label{fig:PrB7Dc}Long term stability due to the automatic alignment control: 
The plot shows the light power transmitted by the
north cavity as a function of time. On the left (time$<7$\,h) the automatic alignment (AA) was
switched off, then after a short period in which the system was prepared
for the C2 run, the cavity is controlled again, this time with the automatic 
alignment turned on so that the power fluctuations are reduced.}
\end{figure}
When the interferometer is at its operating
point the dependence of the quadrant signals on the mirror positions 
can be described by a static matrix. 
For example, if we only consider the vertical alignment of north arm cavity mirrors NI and NE
the signals on the quadrant diode Q71 and Q72 are given as 
\begin{equation}
(Sp_{\rm Q71},Sq_{\rm Q71},Sp_{\rm Q72},Sq_{\rm Q72})= 
{\bf M} \cdot (\Theta_{\rm NI}, \Theta_{\rm NE})
\end{equation}
with $Sp$ representing the in-phase component of the
demodulated signal, $Sq$ the quadrature component, $\Theta_{\rm NI}$ and $\Theta_{\rm NE}$ the
misalignment angles of the cavity mirrors and ${\bf M}$ 
the so-called sensing matrix.

The matrix coefficients depend on the demodulation phase and on the
Gouy phase of the beam on the detector. Electronic phase shifters 
allow to control the demodulation phases such that
the coupling of the various degrees is reduced. 
\mFig{fig:q71tx} shows the
measured and simulated matrix coefficient amplitudes for Q71 as a function of the
demodulation phase. 


The demodulated signal from all 
quadrant diodes are acquired as a digital signal, sampled at 500\,Hz. 
A dedicated processor computes the misalignment angles 
with a linear reconstruction method using the measured sensing matrix. 
This is part of the \emph{Global Control},
a combination of hardware and software which is used to compute 
corrections signals from photo diode error signals \cite{GC}.
 
The Global Control sends the alignment signals  
to digital signal processors (DSP) controlling the suspensions\cite{CITF-top}.
The DSPs apply the necessary control filters to compensate the
mechanical transfer function of the suspended mirror and send the
correction signals to the actuators of the suspension.

\section{Experimental demonstration}
The commissioning of the full \VIRGO detector was started in September 2003 
with the alignment of the IMC output beam through the 3\,km long arms.
Step by step, the interferometer length control systems have been implemented.
In the meantime, the automatic alignment of the
mirrors of the north arm had been implemented. 
In January 2004, the automatic alignment control could be started for the 
first time on one of the long cavities. 


After the length control of the two arm cavities had been implemented,
the interferometer was used in the so-called recombined configuration,
in which also the position of the \bs\ is controlled so that 
the outgoing beams interfere destructively (output port B1).
This operating condition is called \emph{dark fringe}.
It turned out that the longitudinal control is very sensitive
to mis-alignments of the mirrors. A reliable lock
acquisition of the recombined interferometer required the linear
alignment being engaged for the arm cavities.


Approximately every two months a short period of continuous data taking, a 
so-called \emph{commissioning run}, has been scheduled.
The data recorded during these runs are used to monitor the progress of the commissioning of 
the instrument and
allow to evaluate the performance of the subsystems and control systems.
So far, four such runs (C1 to C4) have been performed.

During the C2 run the automatic alignment of the north
arm cavity was used for the first time in a data taking period.
The presence of the alignment control reduces the power fluctuations in the cavity 
considerably and allows long continuous operation without
manual re-alignment of the optics, see \mFig{fig:PrB7Dc}.
Period of continuous operation of up to 32\,hours prove the reliability 
of the system with the alignment control. 

\begin{figure}[h]
\begin{center}
\IG [scale=0.55, angle=0] {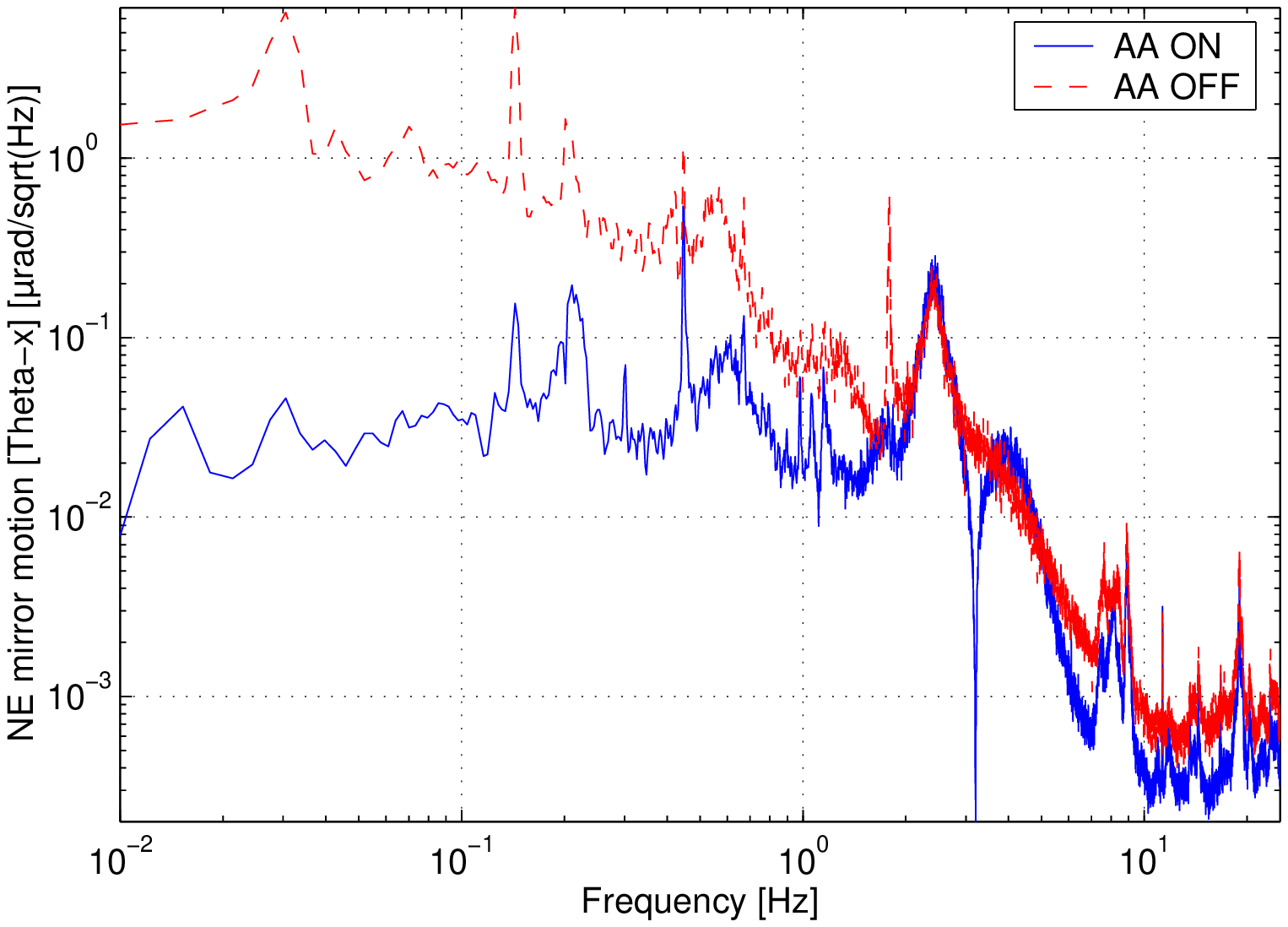}
\end{center}
\caption{\label{fig:NEtx}Comparison
of the noise spectral density of the NE mirror motion in $\Theta_x$ for automatic alignment
switched on and off (measured in-loop). 
Below the unity gain frequency (around 3\,Hz)
the mirror motion is dominated by suspension resonances of the cavity mirrors or the
IMC. For higher frequencies the amplitude falls rapidly to $\approx1$\,nrad$/\sqrt{\rm Hz}$ at 10\,Hz. 
The noise floor for frequencies above 1\,Hz is correlated to the laser frequency noise.}
\end{figure}

In the following commissioning runs
the performance of the system was further improved.
The alignment control systems can be used without change
with one single cavity or within the recombined configuration.
It allows to reduce the RMS alignment fluctuations of the mirrors can 
be reduced to less than $1$\, microradian,
see \mFig{fig:NEtx}. Below the unity gain frequency (between 3\,Hz and 5\,Hz) the
mirror motion is reduced by the control loop. For larger Fourier frequencies a
slight suppression of the alignment in-loop signal can be observed because the
lower RMS misalignment reduces the coupling of noise into the alignment signal.

The system has proved to be easy to use and robust: the automatic
mirror alignment is now part of the standard working
condition of the \VIRGO interferometer. The switching of the
control authority from local controls to linear alignment can be
preformed without losing the longitudinal control. 

The control system is designed to be limited by shot noise at a level of
$10^{-13}$\,radians$/\sqrt{\rm Hz}$. Since the light power in the recombined
configuration is about 500 times lower than in the final configuration 
the shot-noise limit could not yet be achieved. The sensitivity of the 
interferometer is currently too low to be spoiled by 
excess noise from the alignment control loops. 


The linear alignment error signals
could be used to identify resonances of the mirror suspension system
which showed a larger amplitude than expected.
However, in the frequency
range between $100$\,mHz and 10\,Hz the alignment error signals 
as shown in \mFig{fig:NEtx} are coherent with control signal of the
\imc. It is likely that the performance today is limited by the 
beam jitter. Further experiments are currently carried out to
investigate the performance of the input beam control system.



\section{Conclusion}
The \VIRGO interferometer has been operated for several month in the 
recombined configuration, in which the \prc\ is not yet
used. This configuration is not sensitive enough to detect 
gravitational waves but it allowed to implement and test the various subsystems
of the detector. During this commissioning work we have successfully 
implemented for the first time an automatic mirror alignment system for 
a large-scale interferometer using the Anderson technique.
At the same time this demonstrates also for the first time the automatic alignment 
control of a 3\,km long \FP\ with its mirrors suspended by complex \sa.

The delicate optical configuration which is necessary for the
Anderson technique has been set up correctly. The switching 
between the local control systems of the \sa\ and the alignment control
works well. A good long-term stability and noise
performance of the control loops has been demonstrated.
Thus the control design for the linear alignment of the
full detector has been validated.
Currently the implementation of the full automatic alignment
system including the \pom\ is under way.



%


\vfill


\end{document}